\pdfoutput=1

\documentclass[]{spie}  
\usepackage[]{graphicx}
\usepackage{listings}

\title{The Robo-AO automated intelligent queue system} 

\author{Reed L. Riddle\supit{a}, Kristina Hogstrom\supit{a}, Athanasios Papadopoulos\supit{b}, Christoph Baranec\supit{c}, and Nicholas M. Law\supit{d}
\skiplinehalf
\supit{a}Caltech Optical Observatories, California Institute of Technology, 1200 E. California Blvd., MC 11-17, Pasadena, CA, 91125, USA;  
\supit{b}Aristotle University of Thessaloniki, 541 24, Greece;
\supit{c}Institute for Astronomy, University of Hawai$\textquoteleft$i at M\={a}noa, Hilo, HI 96720-2700, USA;  
\supit{d}Department of Physics and Astronomy, University of North Carolina at Chapel Hill, Chapel Hill, NC 27599-3255, USA  
}

\authorinfo{Send correspondence to Reed L. Riddle. E-mail: riddle@caltech.edu, Telephone: 1 626 395 8944\\  
}

\begin{document} 
  \maketitle 

\begin{abstract}

Robo-AO is the first automated laser adaptive optics instrument. In just its second year of scientific operations, it has completed the largest adaptive optics surveys to date, each comprising thousands of targets. Robo-AO uses a fully automated queue scheduling system that selects targets based on criteria entered on a per observing program or per target basis, and includes the ability to coordinate with US Strategic Command automatically to avoid lasing space assets. This enables Robo-AO to select among thousands of targets at a time, and achieve an average observation rate of approximately 20 targets per hour.
\end{abstract}



\section{INTRODUCTION}
\label{sec:intro}  

   \begin{figure}
   \begin{center}
   \begin{tabular}{c}
   \includegraphics[width=15cm]{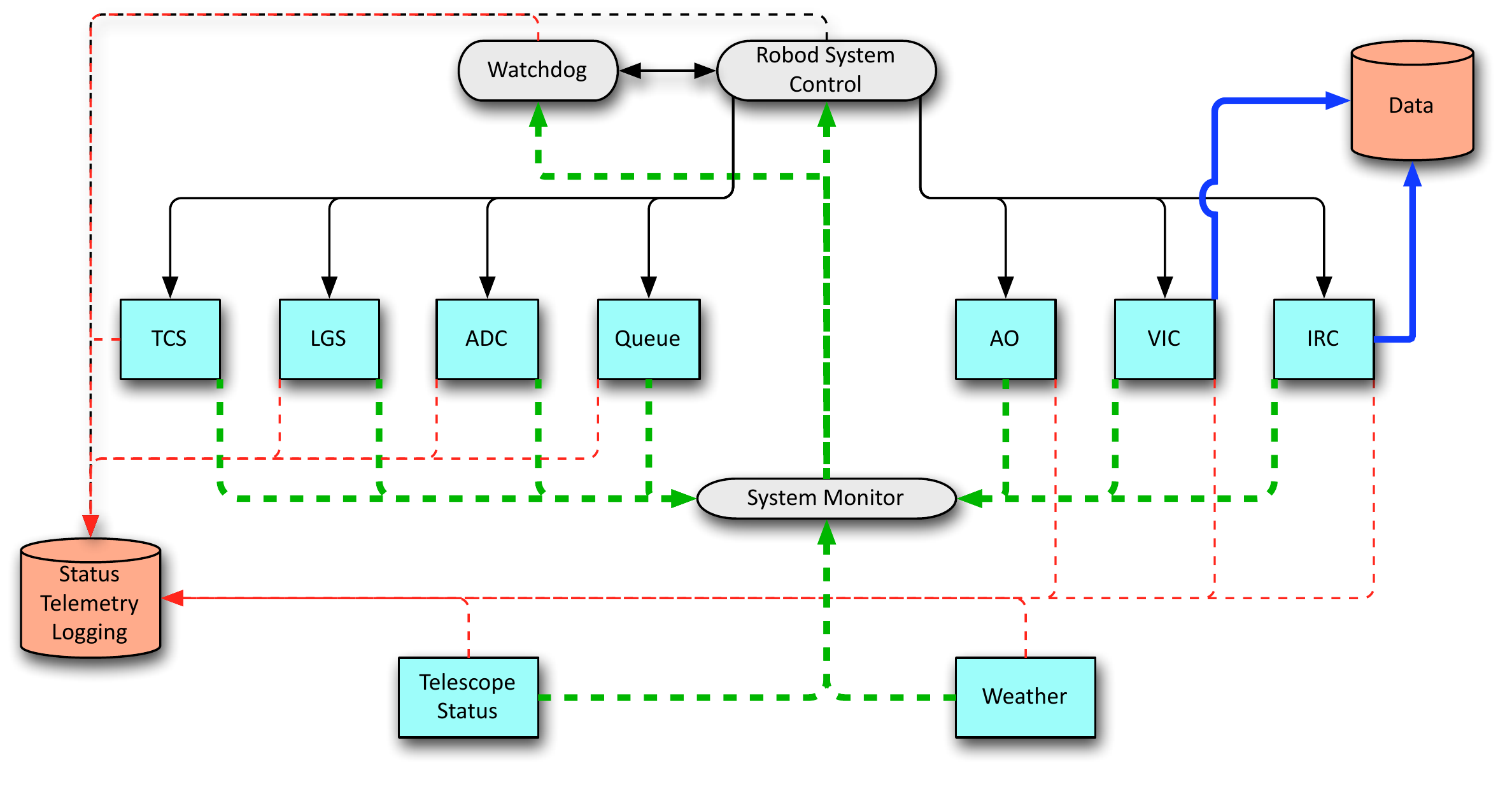}
   \end{tabular}
   \end{center}
   \caption[functions] 
   { \label{fig:functions} The automation software architecture.  Blue boxes are the hardware control subsystem daemons, gray boxes are control or oversight daemons, and red boxes are data file storage.  Red lines with arrows show the paths for telemetry through the operating system, black the command paths, and blue the data paths.}
   \end{figure} 

Robo-AO is an autonomous adaptive optics (AO) instrument that robotically operates a telescope, laser guide star, and science system to observe several different classes of astronomical objects.\cite{Baranec,2013JVE....7250021B,Law}  It is the first system that operates a laser guide star without human oversight, and can operate continuously and robotically through the custom robotic control system software created for the instrument.\cite{2012SPIE.8447E..2OR}
   
The Robo-AO computer uses Fedora 13 as the base operating system. The system does not use a real time kernel; this choice was made to save on complication and increase portability of the software.  In practice, the operation software does not require better than microsecond timing, which is achievable with a non-real-time Linux OS.  All of the source code for the Robo-AO project is written in C++.  At the time of this writing, the software consists of 120,000+ lines of documented source code.  

The Robo-AO control system is composed of several software subsystems.  These subsystems run as daemons in the operating system; each daemon separately manages the operation of hardware and/or software under its control and runs a status monitor to sample subsystem performance and register errors that occur.  Figure~\ref{fig:functions} shows the overall architecture of the entire Robo-AO automated control system.  The subsystem daemons communicate their state through a TCP/IP protocol to a system monitoring service, which is used by the robotic system to control the subsystems and correct for errors.  The robotic system schedules observations and operates the instrumentation to gather data, and a watchdog process monitors the system status and robotic system in case of errors that the robotic system misses or cannot handle.  

The Robo-AO queue system is one of these daemons.  During observations, the robotic sequencing system requests the next target to observe from the queue system, which hands the target, along with the observing parameters, back to the system.  The queue decisions are made based on priorities set in the queue, as well as parameters such as local sidereal time or position of the Moon.  This paper presents the details of the operation of this queue system.

\section{DATA STRUCTURE AND DEFINITIONS}

\begin{figure}
   \begin{center}
   \begin{tabular}{c}
   \includegraphics[height=10cm]{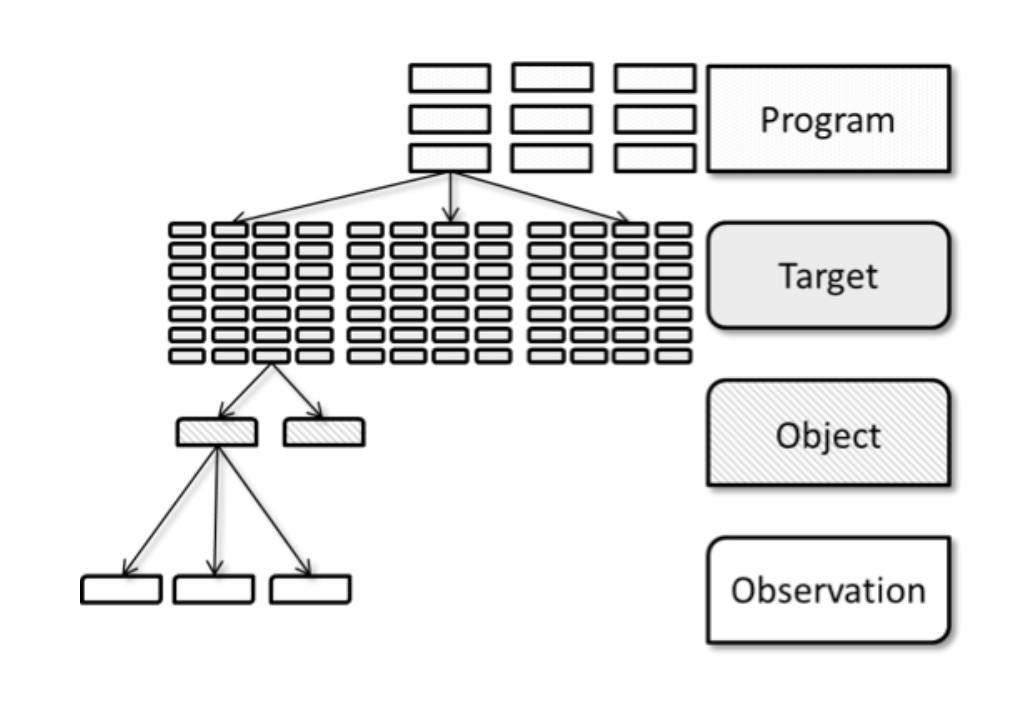}
   \end{tabular}
   \end{center}
   \caption[organization] 
   { \label{fig:organization}Graphical depiction of the queue data organization. }
   \end{figure}

The highest organizational level of the queue system is the program. A program is a complete study of astronomical phenomenon for a specific scientific objective. Programs consist of targets, which are the base units for the queue system. Targets typically contain all the information necessary to complete one sample in the program. A target may consist of only one object, defined as a single point in the sky differentiated by coordinates and magnitude. However, some science goals may require that an object of interest is studied in conjunction with a reference point or that many objects are studied sequentially. In this case, a target would contain more than one object. In addition, an object can have any number of associated observations. Observations can be differentiated by changes in filter wheel, exposure time, camera mode, or many other parameters. Figure~\ref{fig:organization} graphically depicts how the data in the scheduler system are organized.
	
The queue scheduling system is based on a set of XML standard format files, which allows flexibility in the parameters that are included in the file.  The XML files use keywords to determine the required settings and operation parameters for a queue observation.  The files are organized modularly in the scheduler system file directory, and enable complex and diverse observation strategies. There are two different file types:  the programs file, which defines all of the separate observing programs, and the target file, which defines a set of observations for each individual target in a program.  The scheduler system uses four main XML classes for queue data storage: Program, Target, Object, and Observation. Each class contains all variables relevant to that stage of operation of the queue observation.  

Scientists prepare the queue files before the start of a night of observation, and the queue uses the files, and the priorities assigned to the targets and programs, to determine the order of observation.  The queue files are stored in a single directory, with subdirectories for each of the separate programs.  Queue files are updated through the night as observations are completed.

\subsection{Programs XML File}

The definitions for all science programs are contained in one file, Programs.xml, which is located int he top level of the queue files directory.  All programs are grouped together in the file with an example shown below:

\lstset{language=XML}
\begin{lstlisting}
<root>
        <Program>
                <number>0</number>
                <name>High priority</name>
                <person_name>Baranec</person_name>
                <scientific_importance>10</scientific_importance>
                <number_of_targets>81</number_of_targets>
                <counter>77</counter>
                <total_observation_time>16861</total_observation_time>
                <total_science_time>10470</total_science_time>
        </Program>
        <Program>
                <number>1</number>
                <name>Binarity survey</name>
                <person_name>Law</person_name>
                <scientific_importance>2</scientific_importance>
                <number_of_targets>129</number_of_targets>
                <counter>68</counter>
                <total_observation_time>15459</total_observation_time>
                <total_science_time>8040</total_science_time>
	</Program>
        <Program>
                <number>6</number>
                <name>MSOL</name>
                <person_name>Riddle</person_name>
                <scientific_importance>5</scientific_importance>
                <number_of_targets>124</number_of_targets>
                <counter>26</counter>
                <total_observation_time>1360</total_observation_time>
                <total_science_time>540</total_science_time>
        </Program>
</root>
\end{lstlisting}

The modularity of XML files means that, besides the start and end keywords \verb+<root>+ and \verb+<Program>+, the order in which the parameters are listed is irrelevant. The programs themselves do not even have to be entered in the file in sequential order, and parameters can be left out if unnecessary for that program (though in practice all program definitions contain all the keywords).  

Each program is numbered and has a program name and principle investigator attached for informational purposes.  The number of program targets is included, as is a counter that monitors the number of targets that have been completed.  The queue keeps track of the observing time for each program and writes it into the Programs.xml file; multiple TAC allocated programs can be tracked to ensure that they are using their proper time allocation.  The \verb+scientific_importance+ keyword is used to set the priority of the program relative to other programs.  The priority is determined based on scientific and observational programs; for example, a summer program can have its priority turned down during the winter to avoid observations at high airmass.  

The programs file is saved in the top level of the queue data directory. This level also contains a folder for each program which stores all of the target files. By convention, these program folders have the name \verb+Program_#+, with the number corresponding to the number of the program in the programs file.

\subsection{Target XML Files}

The Target XML files are contained in a directory named for their program.  Unlike the programs file, it is not standard to include all of the parameters in a target file. If any unessential parameters are omitted, the system will simply apply default values (which are contained in a configuration file read by the queue system software). Below is an example of a target file, which shows the capability to load multiple objects and observations for a single target. 

\lstset{language=XML}
\begin{lstlisting}
<Target>
        <program_number>0</program_number>
        <number>77</number>
        <name>The coolest star ever</name>
        <visited_times_for_completion>1</visited_times_for_completion>
        <seeing_limit>4</seeing_limit>
        <visited_times>0</visited_times>
        <done>0</done>
        <time_critical>2014-06-25 11:30:00</time_critical>
        <cadence>0</cadence>
        <comment>Hogstrom target</comment>
        <Object>
                <number>1</number>
                <RA>12:34:56.7</RA>
                <dec>+41:31:21.12</dec>
                <epoch>2000</epoch>
                <magnitude>14.5</magnitude>
                <ra_rate>+19.78</ra_rate>
                <dec_rate>-18.70</dec_rate>
                <hour_angle_limit>1.5</hour_angle_limit>
                <done>0</done>
                <Observation>
                        <number>1</number>
                        <exposure_time>90</exposure_time>
                        <ao_flag>1</ao_flag>
                        <filter_code>FILTER_LONGPASS_600</filter_code>
                        <camera_mode>10</camera_mode>
                        <repeat_times>1</repeat_times>
                        <repeated>0</repeated>
                        <done>0</done>
                </Observation>
                <Observation>
                        <number>2</number>
                        <exposure_time>120</exposure_time>
                        <ao_flag>1</ao_flag>
                        <filter_code>FILTER_SLOAN_I</filter_code>
                        <camera_mode>10</camera_mode>
                        <repeat_times>1</repeat_times>
                        <repeated>0</repeated>
                        <done>0</done>
                </Observation>
        </Object>
        <Object>
                <number>2</number>
                <RA>13:23:33.6</RA>
                <dec>+40:30:20.02</dec>
                <epoch>2000</epoch>
                <magnitude>12.5</magnitude>
                <hour_angle_limit>1.5</hour_angle_limit>
                <done>0</done>
                <Observation>
                        <number>1</number>
                        <exposure_time>120</exposure_time>
                        <ao_flag>1</ao_flag>
                        <filter_code>FILTER_LONGPASS_600</filter_code>
                        <camera_mode>10</camera_mode>
                        <repeat_times>1</repeat_times>
                        <repeated>0</repeated>
                        <done>0</done>
                </Observation>
        </Object>
</Target>
\end{lstlisting}

Note that the following units apply:

\begin{itemize}
\item Epoch:  Years
\item Exposure time:  Seconds
\item Moon phase window: Days
\item Sun altitude limit: Degrees
\item Sun distance limit: Degrees
\item Moon distance limit: Degrees
\item Hour angle limit: Hours
\item RA and Dec rates: Arc seconds per hour.
\end{itemize}

Three separate class structures are included in this file.  The \verb+Target+ class lists information about the entire set of observations, including the number of times to visit the target and whether the observation is time critical (i.e. it must be made after a certain time or within a window).  By setting the time critical flag and a cadence, the observer can create a set of periodic observations.  The user can also specify the number of Òvisited times for completionÓ, or how many times the entire target needs to be repeated. Similar to the visited times for completion for the entire target, the user can also choose to repeat specific observations by setting the number of Òtimes to repeatÓ. Finally, the AO flag keyword indicates whether the AO system must be used for the observation.

The \verb+Object+ class structure contains the parameters required to point a telescope at the object.  The queue system has the ability to enter non-sidereal rates in order to observe solar system objects, and an hour angle limit is used to keep observations close to the meridian if necessary.  The stellar magnitude can be used by the robotic system to determine the mode for reentering the telescope (e.g. fainter stars require telescope pointing to a bright library object after a long slew).  Multiple objects can be included in the target queue file.

The \verb+Observation+ class structure hols the information required to set up and acquire scientific data.  Camera mode settings, the filter to use, exposure times, and the number of times a particular observation needs to be repeated are set in this section.  A single object can have multiple observations, and repeat the observation multiple times, which creates a flexible system to achieve whatever observation might be necessary for the scientific program.

	When an observation is made successfully and repeated as necessary, the system  sets the observationÕs ÒdoneÓ flag to 1. The ÒdoneÓ flags for objects and targets are also updated as necessary. If an entire target is complete, then the counter in the program containing the target is incremented. Note that the counter will not increment if only one of several objects in a target have been observed. After every decision, the system rewrites all necessary files in the directory to reflect these changes. In most operating conditions, these files are only loaded once at the beginning of the night, but by rewriting them continuously rather than at the end of the night, the decision data is protected against any interruptions or system failures.  

All queue programs in the system are loaded into a single vector of Program classes in memory. All targets in the system, regardless of their associated program, are stored in one single vector of Target classes, and thus the number of elements in this vector is equal to the sum of the number of targets in each program. Within each instance of a Target class is a vector of Object classes, with one element for each object in that target. Similarly, within each instance of an Object class is a vector of Observation classes, with one element for each observation in that object.

\section{THE DECISION PROCESS}

Once all of the queue data files are loaded into memory, the queue system is ready to determine the next object for the robotic system to observe.  When requested, the queue system runs each of the targets through the decision process, which first eliminates targets that cannot be observed, and then assigns a weight to the remaining targets to determine their priority in the queue at that time.  The optimal object is then chosen, and all observations associated with that object are executed.

\subsection{Elimination Criteria}

	In order to be considered observable, an object must pass a number of requirements. One criterion is simply whether the object is in the sky at the current time. Others are based on the current location of and the objectÕs proximity to the Sun and Moon or other astronomy-based parameters. The user has the option of adjusting the severity of these limits imposed on each object. If the user chooses not to explicitly define a limit, the queue system will assign default values.  In all cases, the object must meet these limits at both the start and end of observations. The location of the Sun and Moon, the moon phase, the airmass, and all necessary inputs are computed using SkyCalc, a suite of astronomical functions developed by John Thorstensen.

\begin{description}
\item[Declination limit:] The object must have a declination viewable to the observatory. 
\item[Hour angle limit:] The object must fall within a certain range of hour angles centered about zero. For instance, the user may require that an object only be observed when it is within one hour from the meridian. The limit in this case would be 1, and the object can only be observed when its hour angle is less than 1 or greater than 23.  
\item[Airmass limit:] The objectÕs associated airmass, tantamount to zenith distance, must be less than the limit. It should be noted that airmass is also used as a weighting criteria, as will be discussed in the following section. However, the user is still given the option of imposing a hard threshold for airmass.
\item[Sun altitude limit:] The Sun must be below a certain altitude in the observatoryÕs horizontal coordinate system. Generally, this limit is used to impose that observations begin after astronomical twilight.
\item[Sun distance limit:] The angular distance between the object and the Sun must be greater than a certain limit. This criterion is similar to Sun altitude in practice, but allows the user more control.
\item[Moon distance limit:] The angular distance between the object and the Moon must be greater than a certain limit.
\item[Moon phase limit:] The moon phase is characterized by the fractional number of days since the last new moon. In specifying a moon phase limit, the user is requiring that the time since the last new moon or to the next new moon is less than the limit. 
\item[Time criticality:]The user has the option of specifying that a target is time-critical, meaning that all objects within the target must be viewed within a specific time window.  If the time-critical flag is set, then the time for observations and also a window in hours around this time must be defined. Since this flag is target-based, the same parameter is applied to all objects within a target. 
\end{description}

\subsubsection{Laser Closure Window}

A unique feature of the Robo-AO queue system is the ability to avoid USSTRATCOM laser closure windows completely.  Most AO systems place closure windows around their observing targets only, as they observe a few to tens of targets per night.  Robo-AO has the capability to observe 1000 targets in less than four nights, and placing single windows around each target is not possible for USSTRATCOM to accomplish.  Instead, we developed a strategy that creates tiles in the sky of roughly equal area, down to a zenith distance of 50 degrees.  Any targets in a tile that is closed to observation are eliminated from consideration by the queue.

About 800 tiles are created in the sky; with so many tiles available, there is always something observable, aside from the rare cases that a space event closes all observations.  Using this system, Robo-AO is able to avoid any closure window conflicts that would cause the system to waste time attempting to observe a target only to find it was not allowed.  It also allows the system to do target of opportunity observations, something that is challenging to achieve with other telescopes that interact with USSTRATCOM.

\subsection{Weighting Criteria}

	The elimination criteria narrow the queue list to only objects that are currently observable. In order to choose the most preferable object, a number of parameters are weighted and the total weights of each object are compared. The user has the option of adjusting the relative importance of each parameter or retaining default values.
	
\begin{description}
\item[Slew time] The queue scheduler must have knowledge of the current telescope and dome positions. First, the total angular distance between the telescope and object and the azimuthal distance between the dome slit and object are both computed. The time required for the telescope slew is estimated by assuming a constant speed. The dome slew time, however, requires a more complex function that accounts for ramp-up and ramp-down rates. Once both slew times are obtained, the longer one is compared to the object duration in order to obtain a weight.
\item[Airmass] The objectÕs airmasses at the start and end of observations are computed. If either of these airmasses exceeds the limit, then the object does not pass the elimination criterion previously discussed and the airmass weight is 0. Otherwise, the starting and ending airmasses are averaged. It is assumed that all objects will have an airmass limit between 1 and 2, which is typical for most astronomical observations. The average airmass of the object is then subtracted from 2 to produce the airmass weight, which is a number between 0 and 1. 

\hspace{3cm}$airmass weight=2-1/2(starting airmass+ending airmass)$

Higher numbers thus correspond to a lower average airmass and are preferable.
\item[Seeing limit] The seeing limit is assigned to a target and all objects within that target have that limit, and is meant to limit observations to a time when seeing conditions are good enough to achieve the required scientific performance.  This isn't currently used in the Robo-AO system as a seeing measurement isn't available.
\item[Scientific priority] Priorities are assigned to specific programs. A high priority and a narrow hour angle limit can be used in conjunction to enforce that objects within a program are only observed when they are near the meridian. In addition, programs with objects that are important, but are in the far north or south sky where observing is not optimal, can be assigned high priorities to ensure that these objects are observed. To this end, it may be useful for the user to split programs into sections based on location in the sky.
\item[Time-critical priority] The time-critical flag was discussed among the elimination criteria, and objects with this flag cannot be viewed outside of the specified time window. However, within that time window, the priority of the object is boosted by a value that increases as time left within the window decreases. In other words, the priority is least at the start of the window and most at the end. The exact function that this boost follows can be changed in a configuration file by the observer.
\end{description}

\section{EXECUTING AN OBSERVATION}

   \begin{figure}
   \begin{center}
   \begin{tabular}{c}
   \includegraphics[height=20.5cm]{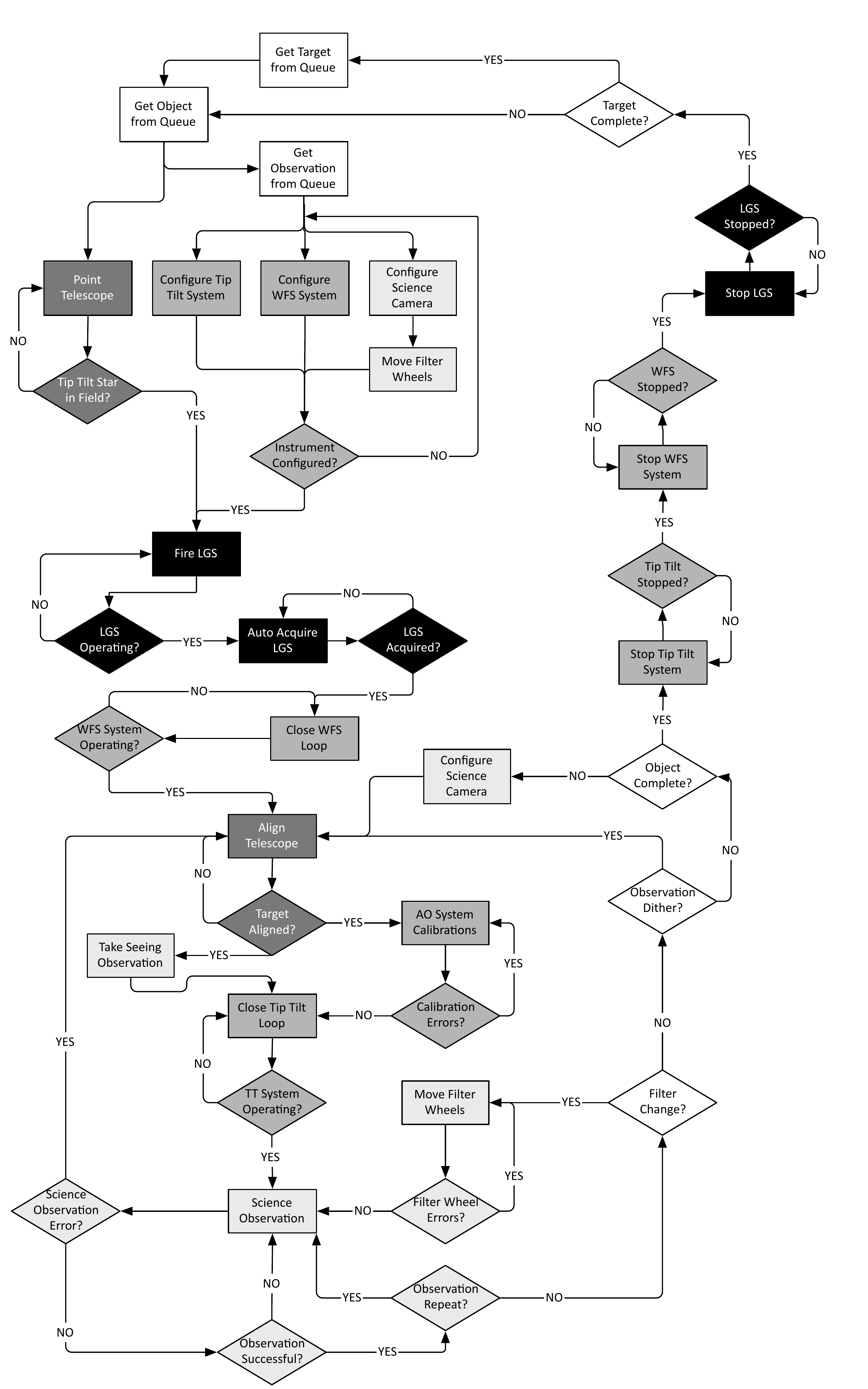}
   \end{tabular}
   \end{center}
   \caption[flowchart] 
   { \label{fig:flowchart} A flow chart demonstrating the operation of the Robo-AO robotic sequencing system.  Note that unshaded steps are controlled by the queue scheduling system.}
   \end{figure}

Figure~\ref{fig:flowchart} shows a flow chart for the operation of the Robo-AO observing system; there are many steps required to propagate the laser, operate the AO system, take science data, and each requires checks to ensure that the system is operating properly.  The queue system determines the parameters that the observing system uses to configure and complete science observations.  Each time the observing system finishes a previous observation target, a new target is requested from the queue system, which calculates the highest priority target and passes its parameters to the observing system.

 \begin{figure}
   \begin{center}
   \begin{tabular}{c}
   \includegraphics[width=15cm]{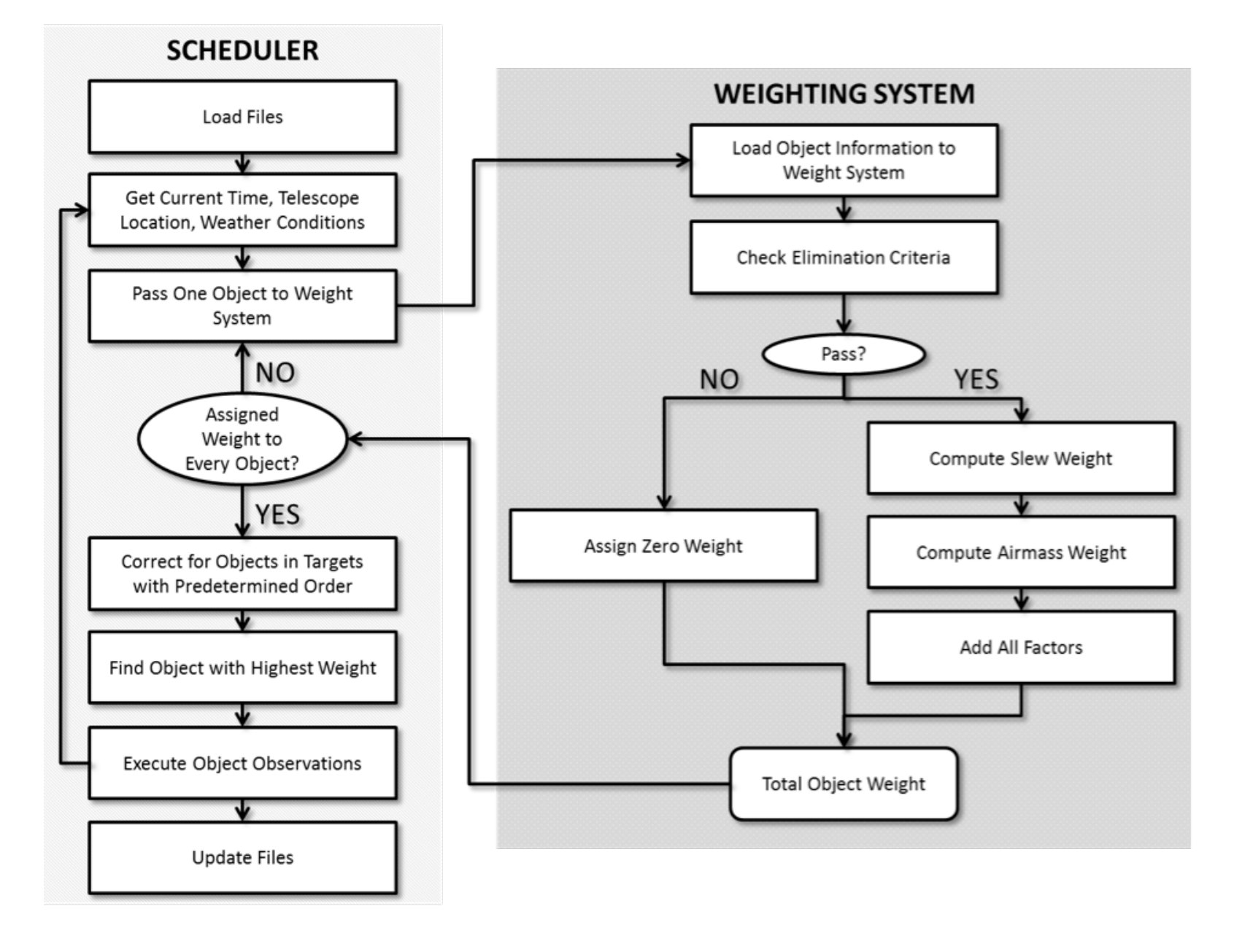}
   \end{tabular}
   \end{center}
   \caption[execution] 
   { \label{fig:execution}The process by which the scheduler makes and executes a decision.}
   \end{figure}

The entire scheduler package consists of two main parts: the scheduler itself and the weighting system. To the scheduler, the weighting system is a black box. The scheduler passes information about a single object, as well as logistics like the current time, current telescope and dome positions, weather conditions, etc. The weighting system then performs all necessary computations to obtain the total weight for the object. The queue software design is modular and allows operators to tweak the weighting system to the needs of a specific observatory. 

Figure~\ref{fig:execution} outlines every task that is performed in the decision-making process and how these tasks are split between the scheduler and the weighting system. First, the scheduler loads the raw data files in the directory. It stores each program as its own structure with all pertaining information as fields. Similarly, each target is stored as its own structure; associated objects are fields of the target, and observations are fields of the objects. The scheduler then gathers information about the observatory, like the current seeing conditions, current time, and current position of the telescope and dome. Next, it passes each object, one by one, to the weighting system. The weighting system first parses this information, and then begins to check if the object passes each elimination criterion. It first checks the criterion that requires the least amount of computing power. If the object passes, then the criterion with the second least amount of computing power is checked, and so on. If the object fails any of the elimination criteria, the system immediately quits and returns a zero weight. In this way, computation time is minimized; calculations are also only done one time and stored between calculation steps (for example, the angular distance to the Moon is calculated a single time).  If the object passes all elimination criteria, then the slew weight and airmass weight are computed, as well as other criteria such as the time critical factor. The system then combines all weighting factors according to the following formula:

\hspace{1.5cm}$total weight=a*slew weight+b*airmass weight+scientific priority+time criticality$

The values of a and b reflect the relative importance of slew weight versus airmass weight, and can be set by the observer in a configuration file. The observer also controls how time criticality is treated in the system.  Once the total weight of the object is computed by the weighting system, it is returned to the scheduler. The scheduler repeats this process until all objects within the system have weights assigned. 

As previously mentioned, the user has the option of specifying whether all objects in a target need to be observed sequentially in a specific order. In this situation, the scheduler must make a correction. The weight of the first object is computed normally. However, for each subsequent object, rather than passing the current telescope position to the weighting system, the scheduler passes the coordinates of the previous object, making the assumption that the telescope would have moved to this point. Thus the actual track of the telescope in observing the full target is simulated. The scheduler then averages the weights from each object and applies this mean weight to every object.

Finally, the scheduler searches for the object with the highest weight. If all objects have a weight of zero, then nothing is observable and the system will wait some set amount of time and try again. This could occur if the scheduler is started before astronomical twilight is over or is still running when twilight begins. If an object is chosen, then the scheduler passes all observation information to the robotic system and waits for a response that the observations were successfully executed. If so, then the observations and the object are marked as completed and the relevant XML files are updated. 

 The amount of time it takes to calculate the queue priority depends on the number of objects loaded into the queue system.  Great care was taken to do the minimum number of calculations when determining the priority for each queue target; this increases the speed and efficiency of the software.  To this point, the maximum number of targets loaded into the queue system has been on the order of 12,000.  Completing the process of determining the priority for this many targets requires about 2.5 seconds.  This is without taking steps such as creating a multithreaded queue that calculates multiple target priorities in parallel; such modifications to the software are planned for the future and will increase its efficiency even further.

\section{CONCLUSION AND FUTURE WORK}

\begin{table}
\small
\begin{center}
\begin{tabular}{p{6.0cm}p{4.0cm}p{1.5cm}cp{1.5cm}}
\hline\hline
\bf{Survey} & \bf{Instrument} & \bf{Method} & \bf{Targets} & \bf{Time} \\
\hline
\raggedright Binarity of the solar neighborhood\cite{Baranec} & P60 (Robo$\,$-AO)  & LGS AO & 3,081 & 172 hours  \\
\raggedright M-Dwarf multiplicity\cite{2012ApJ...754...44J} & \raggedright Calar Alto 2.2-m (AstraLux), NTT (AstraLux) & Lucky & 761 & 300 hours \\
\raggedright Solar-type dwarf multiplicity\cite{Riddle2014} & P60 (Robo$\,$-AO) & LGS AO & 695 & 29 hours  \\
\raggedright Washington Double Star Catalog\cite{2012AJ....143...42H} & SOAR (HRCam) & Speckle, AO+Speckle & 639 & 16 nights  \\
\raggedright Young Solar analogs\cite{2009ApJS..181...62M} & \raggedright KeckII (NIRC2), Hale (PHARO) &  NGS AO & 266 & 47 nights   \\
\raggedright Planets around low-mass stars\cite{2012ApJ...753..142B,Bowler2014} & \raggedright KeckII (NIRC2), Subaru (HiCIAO) & NGS AO &  125 & 12 nights  \\
\raggedright Gemini deep planet survey\cite{2007ApJ...660..770L} & GeminiN (NIRI) & NGS AO & 85 & 84 hours (w.o. overheads)  \\
\raggedright Multiplicity at the bottom of the IMF\cite{2012ApJ...757..141K} & KeckII (NIRC2)  & LGS AO & 78 & 10 nights  \\
\hline
Kepler KOI validation\cite{Law2014a} & P60 (Robo$\,$-AO) & LGS AO & 1,800 & 112 hours  \\
Kepler KOI validation\cite{2012AJ....144...42A} & \raggedright MMT (Aries), Hale (PHARO) & NGS AO & 90 & 12 nights \\
Kepler KOI validation\cite{Lillo-Box} & \raggedright Calar Alto 2.2-m (AstraLux) & Lucky & 98 & 19 nights  \\
\hline\hline
\end{tabular}
\end{center}
\caption{\label{table_1}A representative sample of the largest ground-based diffraction-limited surveys performed with telescopes greater than 1~m in diameter.  }
\end{table}

As of this writing, Robo-AO has completed more than 12,000 queue scheduled robotic observations\cite{Law}.  This has allowed it to complete three of the four largest scientific surveys of high resolution imaging, as shown in Table 1.  The queue system allows Robo-AO to operate very efficiently as a robotic system, and as the most efficient astronomical AO observing system currently in existence.  The Robo-AO configuration time is on average less than 30 seconds, and setting up an adaptive optics system for observation in that amount of time is quite rapid\cite{2006PASP..118..297W}; the next generation large telescopes all will require an automation of tasks of the same order of magnitude as the Robo-AO robotic system in order to achieve their operational requirements\cite{2010SPIE.7736E...2E}.

The queue system continues to become more efficient as improvements are continually made to it and the observing system.  The software itself will be improved with threads for the queue decision process, speeding it up immensely.  The queue daemon itself currently determines the target priority at the time of request, but this can be upgraded so the queue calculation is completed periodically and the most recent results are passed to the robotic system instantly when required.  Both of these upgrades will require analysis for the load they apply to the computer system and effectiveness in queue operation.  

The Robo-AO software, including the queue system, was developed in a modular way to make it easy to replicate the system for other telescopes.  A clone of the current Robo$\,$-AO system is currently being developed for the 2-m IUCAA Girawali Observatory telescope in Maharashtra, India\cite{2002BASI...30..785G}.  In addition to this clone, Pomona College has used the Robo$\,$-AO design and software to develop an AO instrument, built mainly by undergraduate students, that has already achieved on-sky AO correction\cite{Severson}.  The MINERVA project is an array of 0.7m robotic  telescopes that fiber feeds a spectrograph; it will be used for exoplanet research, and the software will be based on the Robo-AO software\cite{2013AAS...22114906H}.  The queue system will be enhanced to enable those observations, increasing the flexibility and capability of the Robo-AO queue system immensely.

Eventually, we plan to release the queue software as a module that other telescopes can use in their software systems, so that other groups looking to develop their own queue system will not have to start from scratch.

\acknowledgments     
 
The Robo-AO system is supported by collaborating partner institutions, the California Institute of Technology and the Inter-University Centre for Astronomy and Astrophysics, and by the National Science Foundation under Grant Nos. AST-0906060, AST-0960343, and AST-1207891, by the Mount Cuba Astronomical Foundation, and by a gift from Samuel Oschin. We are grateful to the Palomar Observatory staff for their ongoing support of Robo-AO on the 60-inch telescope, particularly S. Kunsman, M. Doyle, J. Henning, R. Walters, G. Van Idsinga, B. Baker, K. Dunscombe and D. Roderick.


\bibliography{report}   
\bibliographystyle{spiebib}   

\end{document}